\documentclass[journal]{IEEEtran}
\DeclareUnicodeCharacter{2212}{-}
\ifCLASSINFOpdf
\else
   \usepackage[dvips]{graphicx}
\fi
\usepackage{makecell}
\usepackage{url}
\usepackage{amsfonts}
\usepackage{amsmath}
\usepackage{multirow}
\usepackage{booktabs}
\usepackage{pifont}
\usepackage{cite}

\usepackage[table]{xcolor}  %
\usepackage{footnote}
\usepackage{graphicx}
\usepackage{arydshln}
\usepackage{adjustbox}

\begin{document}

\title{Latent-Level Enhancement with Flow Matching for Robust Automatic Speech Recognition}

\author{Da-Hee Yang, and Joon-Hyuk Chang \IEEEmembership{Senior Member, IEEE}, 

\thanks{D. -H. Yang, and J. -H. Chang (corresponding author) are with the School of
Electronics, Hanyang University, Seoul, 04763, Korea (e-mail: {jchang}@hanyang.ac.kr).}
}

\markboth{Journal of \LaTeX\ Class Files, Vol. 14, No. 8, August 2015}
{Shell \MakeLowercase{\textit{et al.}}: Bare Demo of IEEEtran.cls for IEEE Journals}
\maketitle

\begin{abstract}
Noise-robust automatic speech recognition (ASR) has been commonly addressed by applying speech enhancement (SE) at the waveform level before recognition. However, speech-level enhancement does not always translate into consistent recognition improvements due to residual distortions and mismatches with the latent space of the ASR encoder. In this letter, we introduce a complementary strategy termed latent-level enhancement, where distorted representations are refined during ASR inference. Specifically, we propose a plug-and-play Flow Matching Refinement module (FM-Refiner) that operates on the output latents of a pretrained CTC-based ASR encoder. Trained to map imperfect latents—either directly from noisy inputs or from enhanced-but-imperfect speech—toward their clean counterparts, the FM-Refiner is applied only at inference, without fine-tuning ASR parameters. Experiments show that FM-Refiner consistently reduces word error rate, both when directly applied to noisy inputs and when combined with conventional SE front-ends. These results demonstrate that latent-level refinement via flow matching provides a lightweight and effective complement to existing SE approaches for robust ASR.
\end{abstract}


\IEEEpeerreviewmaketitle
\section{Introduction}
\IEEEPARstart{R}{obust} automatic speech recognition (ASR) under noisy conditions remains a longstanding challenge \cite{robust_asr_spl_2, robustasr1, robustasr5, robust_asr_spl_1, robustasr2, robustasr_add1, robustasr_add2, robustasr_add_yang, robustasr3, robustasr4}. A widely adopted solution is to apply a speech enhancement (SE) front-end, which processes the noisy waveform before recognition \cite{robustasr5, robustasr2}. This strategy is attractive because SE models can be developed independently of the ASR system. However, improved speech quality at the waveform level does not necessarily yield consistent recognition gains. Residual distortions, enhancement artifacts, and mismatches between enhanced speech and the latent space of the ASR encoder often limit reductions in word error rate (WER), leaving a fundamental bottleneck in noise-robust ASR\cite{ASRpro1, ASRpro2, ASRpro3}.

In this letter, we propose a complementary solution that addresses distortion directly at the latent-level of the ASR model. We introduce the Flow Matching Refinement module (FM-Refiner), a generative refinement model that operates on the output representations of a pretrained CTC-based ASR encoder \cite{ctcasr, deepspeech2, e2easr}. The FM-Refiner learns to transform imperfect latents, which may arise from raw noisy inputs or from enhanced speech that still carries residual artifacts, toward distributions characteristic of clean latents. Unlike conventional denoising networks, it leverages flow matching to learn an explicit transport mapping between distorted and clean latent distributions of the ASR encoder. Importantly, this refinement is applied in a plug-and-play manner during inference, without retraining or fine-tuning ASR parameters.

This design offers three advantages. First, it directly improves the representations that matter most for recognition, reducing the mismatch that waveform-level SE alone cannot resolve. Second, it complements existing SE front-ends, enabling a two-stage strategy of waveform enhancement followed by latent refinement. Third, by leveraging flow matching \cite{FM} to learn a deterministic transport mapping, the FM-Refiner provides stable refinement without requiring stochastic sampling \cite{scoreM, flowplc, flowsep, dmellm}.

We evaluate the proposed approach across multiple noise conditions and SE front-ends \cite{convtas, demucs, sgmse+}. Experimental results show that the FM-Refiner consistently reduces WER, whether applied directly on noisy inputs or in combination with SE models. These findings highlight latent-level refinement as a lightweight and effective complement to speech-level enhancement for noise-robust ASR, underscoring the novelty of integrating generative refinement directly into ASR pipelines.

The main contributions of this work are three-fold: (i) we formulate latent-level enhancement for robust ASR, (ii) we design a plug-and-play FM-Refiner that maps imperfect latents toward clean latents without retraining the ASR, and (iii) we provide empirical evidence that latent refinement consistently improves recognition across diverse SE models and noise conditions.


 \begin{figure*}[t]
  \centering
  \includegraphics[width=0.9\linewidth]{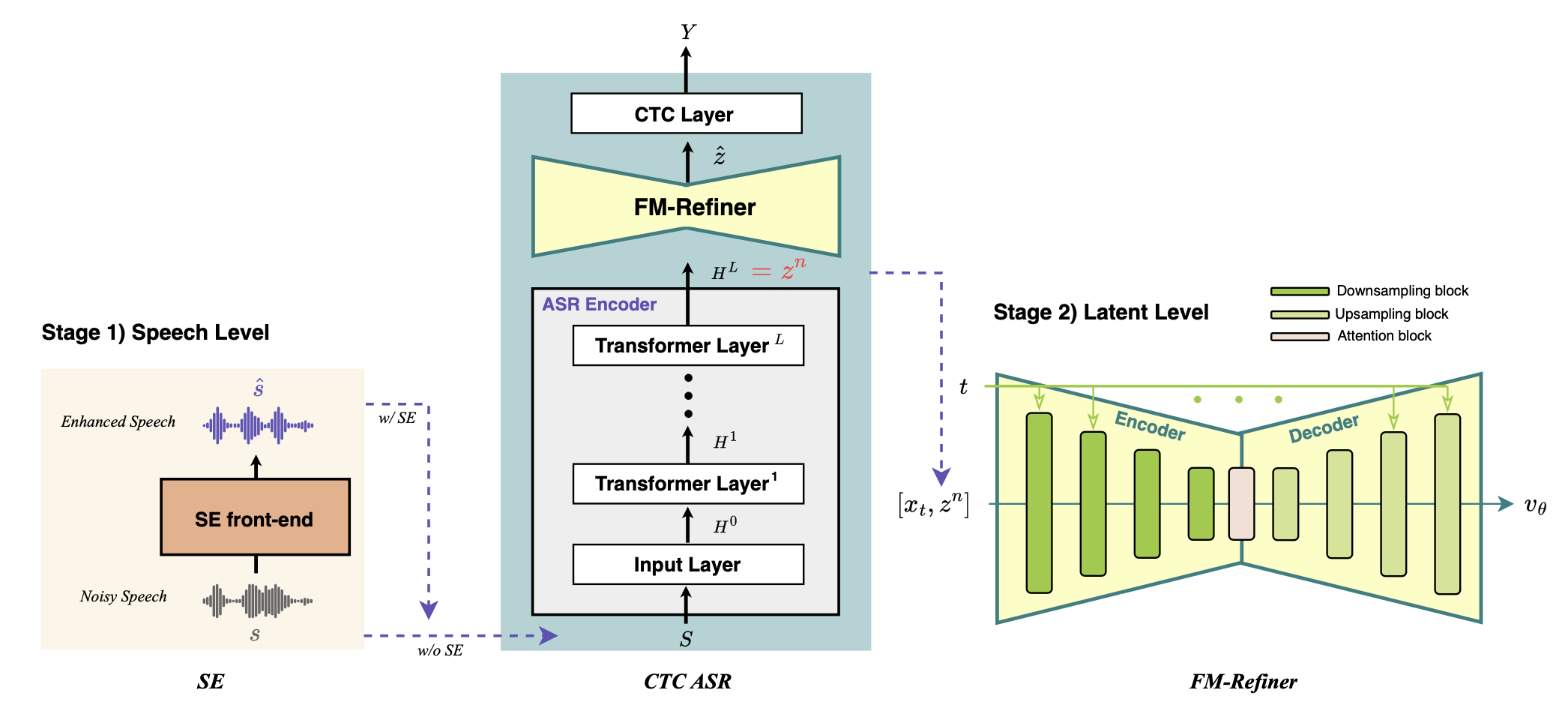}
  \caption{\textbf{Overview of the proposed two-stage framework.} In Stage 1, noisy speech $s$ is optionally enhanced by an SE front-end, producing an enhanced speech for the pretrained CTC-based ASR encoder. In Stage 2, the encoder output latents $z^{n}$ are refined by the FM-Refiner, which produces improved representations $\hat{z}$ for the CTC layer. All modules (SE, ASR, FM-Refiner) are pretrained separately, and the figure illustrates the decoding process where the FM-Refiner is applied in a plug-and-play manner. The FM-Refiner is implemented with a U-Net architecture trained via flow matching.}
  \label{fig:main}
\end{figure*}

\section{Modeling Approach}
\subsection{Framework Overview}
Fig. \ref{fig:main} illustrates the overall framework of the proposed method. In a conventional pipeline for noise-robust ASR, an SE front-end is applied to suppress noise at the waveform level, and the resulting enhanced speech is then passed into a pretrained ASR model. In this work, we assume that the ASR model remains fixed without task-specific retraining, as retraining with new noise conditions or domain-specific data is often impractical. While SE can improve perceptual speech quality, residual artifacts and mismatches with the ASR encoder feature space often persist, which limits recognition performance. 

To address this limitation, we introduce an additional refinement stage that operates at the latent-level of the ASR. Specifically, after the SE front-end produces an enhanced waveform, the signal is passed through the pretrained ASR encoder. Its output latent representation, $\boldsymbol{H}^{L}$, is then refined by the FM-Refiner. The FM-Refiner is pretrained separately to map noisy or enhanced latents toward clean latents, and can be inserted in a plug-and-play manner, without requiring any fine-tuning of the ASR model.
The refined latents, $\hat{z}$, are finally fed into the CTC layer to produce text outputs. This architecture integrates speech-level and latent-level processing: SE mitigates noise in the waveform domain, and FM-Refiner further corrects residual mismatches in the latent space of the ASR. 
By combining the two stages, the framework produces CTC inputs that are better aligned with the classifier’s learned decision space.

\subsection{CTC-based ASR Architecture}
The ASR model employed in this work is based on the connectionist temporal classification (CTC) framework \cite{ctcasr}. The CTC-based ASR system is designed to predict a target transcription sequence $\boldsymbol{Y}$ from an input speech sequence $\boldsymbol{S}$ of length $T$, without requiring explicit frame-level alignment between inputs and outputs.
The encoder of the ASR model consists of 12 transformer layers, each composed of multi-head self-attention, feed-forward networks, residual connections, and layer normalization \cite{transformer}. Given the acoustic feature sequence $\boldsymbol{S}$($\hat{\boldsymbol{S}}$) extracted from the noisy(enhanced) waveform, the encoder maps it into a higher-level latent representation:
\begin{equation}
\begin{aligned}
\boldsymbol{H}^{0} &= \textrm{InputLayer}(\boldsymbol{S}\ (\hat{\boldsymbol{S}})), \\
\boldsymbol{H}^{l} &= \textrm{TransformerLayer}^{l}(\boldsymbol{H}^{l-1}), \quad (1 \leq l \leq L)
\end{aligned}
\end{equation}
where $\boldsymbol{H}^{l}$ denotes the output representation of the $l$-th transformer layer.
The final latent representation $\boldsymbol{H}^{12}$, where $L=12$, is fed into a linear projection layer followed by a softmax to compute the frame-wise posterior distribution over subword units. 
The CTC loss is defined as the negative log-likelihood of the target sequence $\boldsymbol{Y}$ given the input sequence $\boldsymbol{S}$, marginalized over all possible alignment paths $B(\boldsymbol{Y},\boldsymbol{S})$, where each alignment $\hat{\boldsymbol{y}}$ has length $N$:
\begin{equation}
P_{CTC}(\boldsymbol{Y}|\boldsymbol{S}) = \sum_{\hat{\boldsymbol{y}} \in B(\boldsymbol{Y}, \boldsymbol{S})} \prod_{n=1}^{N} P(\hat{\boldsymbol{y}}_{n}|\boldsymbol{S}),
\end{equation}
\begin{equation}
\mathcal{L}_{CTC} = - \ln P_{CTC}(\boldsymbol{Y}|\boldsymbol{S}).
\end{equation}
This formulation allows the ASR system to learn from encoder outputs without requiring explicit frame-to-label alignments, making the latent space directly compatible with the proposed FM-Refiner.

\subsection{FM-Refiner}
\subsubsection{Flow Matching Preliminary}
Flow matching (FM) \cite{FM} is a generative modeling framework that learns a continuous transformation between a prior distribution and a target data distribution by estimating the underlying vector field that governs the evolution of intermediate samples. 
Formally, let $x_0 \sim q_0(x)$ denote a sample from an initial distribution and $x_1 \sim q_1(x)$ denote a target sample from the data distribution. 
FM defines a time-dependent mapping $f(x_0, t) = x_t$, where $t \in [0, 1]$ represents the continuous interpolation between $x_0$ and $x_1$. 
The dynamics of this mapping follow an ordinary differential equation (ODE):
\begin{equation}
    \frac{df}{dt} = \frac{dx_t}{dt} = u_t,
\end{equation}
where $u_t$ is the vector field describing the instantaneous direction of transport from $x_0$ to $x_1$ at time $t$.

In practice, the exact $u_t$ is unknown and is instead approximated by a neural network $v_\theta(x_t, t)$. 
The FM objective trains $v_\theta$ to minimize the discrepancy between the true and predicted vector fields:
\begin{equation}
    \mathcal{L}_{\text{FM}} = \mathbb{E}_{x_0, x_1, t} \left[ \| v_\theta(x_t, t) - u_t \|_2^2 \right].
\end{equation}
To make the training objective tractable, an affine probability path between $x_0$ and $x_1$ is introduced. 
A common choice is the optimal transport (OT) path, which interpolates the two endpoints linearly with a noise-controlling factor $\sigma_{\min}$:
\begin{equation}
    f(x_0, t) = t x_1 + \left( 1 - (1 - \sigma_{\min})t \right) x_0.
\end{equation}
The corresponding vector field is then given by
\begin{equation}
    u_t = x_1 - (1 - \sigma_{\min}) x_0,
\end{equation}
which is independent of $t$ and therefore enables efficient and stable learning.

Once trained, the FM model generates samples by integrating the learned vector field $v_\theta$ over time using an ODE solver. 
Given an initial sample $x_0 \sim q_0(x)$, the solver iteratively updates $x_t$ according to
\begin{equation}
    x_{t + \Delta t} = x_t + \Delta t \cdot v_\theta(x_t, t),
\end{equation}
where $\Delta t = 1/N$ and $N$ is the number of sampling steps.  \\

\subsubsection{FM-Refiner for ASR Latent Representation}
Building upon this foundation, we introduce the FM-Refiner, a refinement module that leverages flow matching to enhance ASR latent representations. 
The key idea is to align imperfect ASR latents with their clean counterparts through the vector field estimated by FM network, thereby improving recognition robustness under noisy conditions.

Specifically, given a noisy input utterance $\mathbf{s}$, we first obtain its latent representation using a pretrained ASR encoder. 
Let $\mathbf{z}^n$ and $\mathbf{z}^c$ denote the noisy and clean ASR latents, respectively. 
During training, $\mathbf{z}^n$ serves as the starting distribution $x_0$, while $\mathbf{z}^c$ provides the target $x_1$. 
Intermediate states $x_t$ are then constructed along the OT path between $\mathbf{z}^n$ and $\mathbf{z}^c$, and the FM-Refiner is trained to predict the corresponding vector field.

The FM-Refiner adopts a multi-resolution U-Net backbone \cite{song2020score} tailored for sequential latent features. 
At each time step, the diffused latent $x_t$ is concatenated with the noisy latent $\mathbf{z}^n$ and fed into the encoder as shown in Fig \ref{fig:main}. 
The encoder progressively reduces the sequence resolution via residual convolutional blocks with group normalization and SiLU activation \cite{silu}, while the decoder reconstructs the resolution using upsampling layers and skip connections to retain fine-grained detail. 
A global attention bottleneck is incorporated to capture contextual dependencies across the sequence. 
In addition, time step information $t$ is embedded via positional embeddings and injected into each residual block, enabling the network to modulate its feature transformations over the probability path.

The training objective follows the conditional flow matching loss:
\begin{equation}
    \mathcal{L}_{\text{FM-Refiner}} = \mathbb{E}_{\mathbf{z}^n, \mathbf{z}^c, t} \left\| v_\theta(x_t, \mathbf{z}^n, t) - \left(\mathbf{z}^c - (1-\sigma_{\min}) \mathbf{z}^n \right) \right\|_2^2.
\end{equation}
Here, the conditioning on $\mathbf{z}^n$ ensures that the refinement process remains anchored to the input signal, preventing over-smoothing or loss of speaker-specific information.

At inference, only the noisy latent $\mathbf{z}^n$ is available. 
The FM-Refiner initializes from $\mathbf{z}^n$ and iteratively integrates the learned vector field to generate the refined latent $\hat{\mathbf{z}}$, which is then passed to the CTC layer. 
This refinement process effectively bridges the gap between noisy and clean representations, leading to improved word recognition accuracy across a wide range of noise conditions. When an SE front-end is applied, the noisy latent $\mathbf{z}^n$ is replaced with the enhanced latent representation, which then serves as the initialization for the refinement process.

\section{Experimental Setup}
\subsection{Datasets}
The ASR model was trained on the Wall Street Journal (WSJ) corpus \cite{wsjdata} using standard data augmentation techniques, without fine-tuning on noisy speech.
To train the SE front-ends, paired clean and noisy utterances were generated by mixing WSJ training and validation utterances with randomly selected CHiME-4 noise at SNRs uniformly sampled from -5 to 10 dB. 
The CHiME-4 dataset includes recordings from street, café, bus, and pedestrian environments, providing diverse and realistic noise conditions \cite{chime}. 
For evaluation, WSJ test utterances were mixed with CHiME-4 noise at SNRs from $\{-5, -2, 0, 2, 5, 10\}$ dB. 
WER was measured under two conditions: (i) direct decoding of noisy speech and (ii) decoding of enhanced speech after passing through the SE front-end. 
 
\begin{table*}[!t] \ref{tab1}
    \caption{Experimental results of unprocessed input and various SE front-ends, with and without the proposed FM-Refiner. The FM-Refiner consistently improves WER (lower is better) across all conditions. SE performance is reported for reference  and is not directly related to the proposed method.}
  \label{tab1}
  \centering
  \resizebox{0.9\linewidth}{!}{
  \definecolor{rose}{RGB}{186, 110, 126}  
  \definecolor{skycomp1}{RGB}{143,170,188}  
  \definecolor{skycomp2}{RGB}{120,155,180}  
  \definecolor{green1}{RGB}{185, 210, 155}  
  \begin{tabular}{cccccccccc}
    \toprule
\multicolumn{2}{c}{\multirow{3}{*}{Method}} & \multicolumn{7}{c}{ASR performance}     & SE performance                 \\ \addlinespace[0.5mm] \cline{3-10} \addlinespace[0.5mm]
\multicolumn{2}{c}{}                       & \multicolumn{7}{c}{WER($\downarrow$)} & DNSMOS($\uparrow$)                   \\ \addlinespace[0.5mm]\cline{3-10} \addlinespace[0.5mm]
\multicolumn{2}{c}{}                        & -5 dB & -2 dB & 0 dB & 2 dB & 5 dB & 10 dB & \textbf{Avg.} & [-5, 10] dB  \\ \addlinespace[0.5mm]\hline \addlinespace[0.8mm]
Unprocessed          &                      & 85.6  & 79.7  & 73.4 & 63.9 & 51.7 & 39.3 & 65.60 & 2.74  \\ \addlinespace[0.5mm]
                     & w/ FM-Refiner        & \cellcolor{green1!40}80.9  & \cellcolor{green1!40}73.5  & \cellcolor{green1!40}65.9 & \cellcolor{green1!40}56.4 & \cellcolor{green1!40}45.6 & \cellcolor{green1!40}35.2 & \cellcolor{green1!40}\textbf{59.58} & -  \\ \addlinespace[0.5mm] \hline \addlinespace[0.5mm]
\textit{SE front-ends}         &                      &       &       &      &      &      &   &   \\ \addlinespace[0.8mm]
Conv-TasNet          &                      & 54.2  & 44.8  & 39.7 & 35.1 & 31.0 & 27.7 & 38.75 & 3.73  \\ \addlinespace[0.5mm]
                     & w/ FM-Refiner        & \cellcolor{green1!40}50.2  & \cellcolor{green1!40}40.9  & \cellcolor{green1!40}36.9 & \cellcolor{green1!40}33.0 & \cellcolor{green1!40}28.8 & \cellcolor{green1!40}26.8 & \cellcolor{green1!40}\textbf{36.10}& -  \\ \addlinespace[0.5mm] \cdashline{2-10} \addlinespace[0.5mm]
DEMUCS               &                      & 60.9  & 49.2  & 43.7 & 38.0 & 31.7 & 27.2 & 41.78 & 3.73   \\ \addlinespace[0.5mm]
                     & w/ FM-Refiner        & \cellcolor{green1!40}55.9  & \cellcolor{green1!40}46.3  & \cellcolor{green1!40}42.3 & \cellcolor{green1!40}35.8 & \cellcolor{green1!40}30.2 & \cellcolor{green1!40}26.7 &
                     \cellcolor{green1!40}\textbf{39.53} &-  \\ \addlinespace[0.5mm] \cdashline{2-10} \addlinespace[0.5mm]
SGMSE+               &                      & 56.2  & 45.9  & 40.8 & 36.1 & 31.3 & 27.0 & 39.55 & 3.98  \\ \addlinespace[0.5mm]
                     & w/ FM-Refiner        & \cellcolor{green1!40}54.2  & \cellcolor{green1!40}43.7  & \cellcolor{green1!40}39.1 & \cellcolor{green1!40}34.4 & \cellcolor{green1!40}29.9 & \cellcolor{green1!40}26.5 &
                     \cellcolor{green1!40}\textbf{37.97} & -   \\ 
    \bottomrule
  \end{tabular}}
\end{table*}

\subsection{Model Configuration} 
\begin{itemize}
    \item CTC-based ASR \\
    We employed a CTC-based ASR model built with the ESPnet toolkit \cite{espnet, espnetgit}. The input features were 80-dimensional log Mel-filterbank coefficients with a 25 ms window size and 10 ms frame shift. The encoder consisted of 12 Transformer layers, each equipped with 4 attention heads and a hidden dimension of 256. The position-wise feed-forward networks had an inner dimension of 2048. A CTC layer followed the encoder to predict the posterior distributions over the output label set. The model was trained for 100 epochs with a batch size of 32. The Adam optimizer \cite{adam} was used, along with SpecAugment for data augmentation and checkpoint averaging to stabilize the final model parameters. \\

    \item FM-Refiner\\
    We employed a U-Net backbone to implement the proposed FM-Refiner. The network consists of four downsampling layers and four upsampling layers, where the ASR encoder outputs serve as the input latent representations. Specifically, the ASR encoder produces time-sequence features with a hidden dimension of 256, which are progressively refined through the flow matching process. 
    The model was trained using Adam optimizer with a batch size of 16 for 200 epochs on a single NVIDIA A100 GPU.
    During inference, the FM-Refiner was executed with three sampling steps. Detailed architectural parameters beyond the number of layers follow the implementation provided in the official code\footnote{\url{https://github.com/sp-uhh/sgmse.git}}.
\end{itemize}

\subsection{SE front-end models} 
To demonstrate the generality of the proposed FM-Refiner, we employed several SE front-end models, including both discriminative and generative approaches. This setup allows us to evaluate the effectiveness of latent-level refinement across a wide range of enhancement techniques, ensuring that the proposed method is not tied to a specific SE architecture or objective.

\begin{itemize}
    \item Conv-TasNet \cite{convtas}: A time-domain discriminative SE model based on stacked temporal convolution networks that predicts a time-varying mask for the input waveform and applies it to produce an enhanced waveform.
    
    \item DEMUCS \cite{demucs}: A waveform-based discriminative SE model built on a U-Net architecture. It maps noisy waveforms to clean waveforms through an encoder-decoder structure.
    
    \item SGMSE+ \cite{sgmse+}: A generative SE model that operates in the spectrogram domain. It uses a noise-conditional score network to model the clean speech distribution, enabling sample refinement through iterative denoising.
\end{itemize}

\section{Experimental Results}
Table~\ref{tab1} presents the WER performance of various SE models under different signal-to-noise ratio (SNR) conditions, both with and without the proposed FM-Refiner. Several key observations can be drawn from these results. First, even when applied directly to noisy speech without any SE front-end, the FM-Refiner substantially improved ASR accuracy across all SNR levels, ranging from the highly adverse condition of -5 dB to the relatively clean 10 dB environment. This indicates that the pretrained FM-Refiner can refine ASR latent representations in a plug-and-play manner without retraining the ASR encoder.

Second, when combined with conventional SE front-ends such as Conv-TasNet, DEMUCS, and SGMSE+, the FM-Refiner consistently reduced WER, demonstrating its effectiveness as a complementary refinement stage. Notably, the performance gains were observed uniformly across all tested conditions, suggesting that the FM-Refiner is robust to different SE architectures and noise severities. This consistency highlights the flexibility of the proposed approach in integrating with existing SE pipelines.

An interesting observation arises when comparing different SE front-ends. We evaluated SE quality using DNSMOS \cite{dnsmos}, a perceptual metric that reflects the perceived naturalness of speech. As shown in Table~\ref{tab1}, the generative model SGMSE+ achieved the highest DNSMOS score, indicating more natural-sounding enhanced speech. However, in terms of downstream ASR, its WER performance lagged behind the discriminative model Conv-TasNet. 
This discrepancy highlights that perceptual or signal-level improvements do not necessarily yield better ASR accuracy, showing that waveform-level enhancement alone is insufficient.
In contrast, the FM-Refiner directly refines the ASR latent space, bridging the gap between noisy and clean representations and yielding improvements that align more closely with recognition objectives.

Overall, the results demonstrated that FM-Refiner offers a systematic and effective solution for refining ASR latent representations. Its consistent improvements across both unprocessed and SE-processed inputs validated the necessity of latent-level enhancement in complementing traditional SE methods.

\section{Conclusion}
We proposed FM-Refiner, a latent-level refinement framework designed to enhance the robustness of CTC-based ASR under noisy conditions. By applying a flow matching module to ASR encoder outputs, FM-Refiner progressively transformed noisy latents into cleaner counterparts, serving as a lightweight and effective refinement step. Experiments with multiple SE front-ends showed consistent improvements in word recognition accuracy across diverse noise levels, highlighting its flexibility and applicability. These results indicate that incorporating probabilistic refinement into ASR latents offers a novel way to bridge the gap between noisy and clean representations.

\bibliographystyle{ieeetr}
\bibliography{strings}

@article{convtas,
  title={\textup{Conv-TasNet}: Surpassing ideal time--frequency magnitude masking for speech separation},
  author={Luo, Yi and Mesgarani, Nima},
  journal={IEEE/ACM Transactions on Audio, Speech, and Language Processing},
  volume={27},
  number={8},
  pages={1256--1266},
  year={2019},
  publisher={IEEE}
}

@article{sgmse+,
  title={Speech enhancement and dereverberation with diffusion-based generative models},
  author={Richter, Julius and Welker, Simon and Lemercier, Jean-Marie and Lay, Bunlong and Gerkmann, Timo},
  journal={IEEE/ACM Transactions on Audio, Speech, and Language Processing},
    volume={31},
    number={},
    pages={2351--2364},
    year={2023},
  publisher={IEEE}
}

@inproceedings{demucs,
  title={Real time speech enhancement in the waveform domain},
  author={Defossez, Alexandre and Synnaeve, Gabriel and Adi, Yossi},
  booktitle={Proc. INTERSPEECH},
  pages={3291--3295},
  year={2020}
}

@inproceedings{ctcasr,
  title={Towards end-to-end speech recognition with recurrent neural networks},
  author={Graves, Alex and Jaitly, Navdeep},
  booktitle={Proc. International Conference on Machine Learning \textup{(ICML)}},
  pages={1764--1772},
  year={2014}
}

@inproceedings{deepspeech2,
  title={Deep speech 2: End-to-end speech recognition in english and mandarin},
  author={Amodei, Dario and Ananthanarayanan, Sundaram and Anubhai, Rishita and Bai, Jingliang and Battenberg, Eric and Case, Carl and Casper, Jared and Catanzaro, Bryan and Cheng, Qiang and Chen, Guoliang and others},
  booktitle={Proc. International Conference on Machine Learning (ICML)},
  pages={173--182},
  year={2016}
}

@inproceedings{e2easr,
  title={State-of-the-art speech recognition with sequence-to-sequence models},
  author={Chiu, Chung-Cheng and Sainath, Tara N and Wu, Yonghui and Prabhavalkar, Rohit and Nguyen, Patrick and Chen, Zhifeng and Kannan, Anjuli and Weiss, Ron J and Rao, Kanishka and Gonina, Ekaterina and others},
  booktitle={Proc. IEEE International Conference on Acoustics, Speech and Signal Processing (ICASSP)},
  pages={4774--4778},
  year={2018}
}

@inproceedings{transformer,
  title={Attention is all you need},
  author={Vaswani, Ashish and Shazeer, Noam and Parmar, Niki and Uszkoreit, Jakob and Jones, Llion and Gomez, Aidan N and Kaiser, {\L}ukasz and Polosukhin, Illia},
  booktitle={Proc. Advances in Neural Information Processing Systems \textup{(NeurIPS)}},
  volume={30},
  year={2017}
}

@article{ASRpro1,
  title={Bridging the gap between monaural speech enhancement and recognition with distortion-independent acoustic modeling},
  author={Wang, Peidong and Tan, Ke and Wang, DeLiang},
  journal={IEEE/ACM Transactions on Audio, Speech, and Language Processing},
  volume={28},
  pages={39--48},
  year={2019},
  publisher={IEEE}
}

@inproceedings{ASRpro2,
  title={One-Pass Single-Channel Noisy Speech Recognition Using a Combination of Noisy and Enhanced Features.},
  author={Fujimoto, Masakiyo and Kawai, Hisashi},
  booktitle={Proc. INTERSPEECH},
  pages={486--490},
  year={2019}
}

@article{ASRpro3,
  title={How Bad Are Artifacts?: Analyzing the Impact of Speech Enhancement Errors on ASR},
  author={Iwamoto, Kazuma and Ochiai, Tsubasa and Delcroix, Marc and Ikeshita, Rintaro and Sato, Hiroshi and Araki, Shoko and Katagiri, Shigeru},
  journal={arXiv preprint arXiv:2201.06685},
  year={2022}
}

@inproceedings{song2020score,
  title={Score-based generative modeling through stochastic differential equations},
  author={Song, Yang and Sohl-Dickstein, Jascha and Kingma, Diederik P and Kumar, Abhishek and Ermon, Stefano and Poole, Ben},
  booktitle={Proc. International Conference on Learning Representations \textup{(ICLR)}},
  year={2021}
}

@inproceedings{espnet,
  title={Espnet: End-to-end speech processing toolkit},
  author={Watanabe, Shinji and others},
  booktitle={Proc. INTERSPEECH},
  year={2018}
}

@article{espnetgit,
  title={End-to-End Speech Processing Toolkit},
  journal={https://github.com/espnet/espnet},
  year={cited March 21 2018.}
}

@article{robust_asr_spl_2,
  title={Robust speech recognition by nonlocal means denoising processing},
  author={Xu, Haitian and Tan, Zheng-Hua and Dalsgaard, Paul and Lindberg, Brge},
  journal={IEEE signal processing letters},
  volume={15},
  pages={701--704},
  year={2008},
  publisher={IEEE}
}

@article{robustasr1,
  title={An overview of noise-robust automatic speech recognition},
  author={Li, Jinyu and Deng, Li and Gong, Yifan and Haeb-Umbach, Reinhold},
  journal={IEEE/ACM Transactions on Audio, Speech, and Language Processing},
  volume={22},
  number={4},
  pages={745--777},
  year={2014},
  publisher={IEEE}
}

@inproceedings{robustasr5,
  title={Speech enhancement with {LSTM} recurrent neural networks and its application to noise-robust {ASR}},
  author={Weninger, Felix and Erdogan, Hakan and Watanabe, Shinji and Vincent, Emmanuel and Le Roux, Jonathan and Hershey, John R and Schuller, Bj{\"o}rn},
  booktitle={Proc. International Conference on Latent Variable Analysis and Signal Separation \textup{(LCA/ICA)}},
  pages={91--99},
  year={2015},
  organization={Springer}
}

@article{robust_asr_spl_1,
  title={DNN uncertainty propagation using GMM-derived uncertainty features for noise robust ASR},
  author={Nathwani, Karan and Vincent, Emmanuel and Illina, Irina},
  journal={IEEE Signal Processing Letters},
  volume={25},
  number={3},
  pages={338--342},
  year={2018},
  publisher={IEEE}
}

@article{robustasr2,
  title={Complex spectral mapping for single-and multi-channel speech enhancement and robust {ASR}},
  author={Wang, Zhong-Qiu and Wang, Peidong and Wang, DeLiang},
  journal={IEEE/ACM Transactions on Audio, Speech, and Language Processing},
  volume={28},
  pages={1778--1787},
  year={2020}
}

@article{robustasr_add1,
  title={Gated recurrent fusion with joint training framework for robust end-to-end speech recognition},
  author={Fan, Cunhang and Yi, Jiangyan and Tao, Jianhua and Tian, Zhengkun and Liu, Bin and Wen, Zhengqi},
  journal={IEEE/ACM Transactions on Audio, Speech, and Language Processing},
  volume={29},
  pages={198--209},
  year={2020}
}

@inproceedings{robustasr_add2,
  title={Interactive feature fusion for end-to-end noise-robust speech recognition},
  author={Hu, Yuchen and Hou, Nana and Chen, Chen and Chng, Eng Siong},
  booktitle={Proc. IEEE International Conference on Acoustics, Speech and Signal Processing \textup{(ICASSP)}},
  pages={6292--6296},
  year={2022}
}

@inproceedings{robustasr_add_yang,
  title={FiLM Conditioning with Enhanced Feature to the Transformer-based End-to-End Noisy Speech Recognition.},
  author={Yang, Da-Hee and Chang, Joon-Hyuk},
  booktitle={Proc. INTERSPEECH},
  pages={4098--4102},
  year={2022}
}

@inproceedings{robustasr3,
  title={Improving noise robust automatic speech recognition with single-channel time-domain enhancement network},
  author={Kinoshita, Keisuke and Ochiai, Tsubasa and Delcroix, Marc and Nakatani, Tomohiro},
  booktitle={Proc. IEEE International Conference on Acoustics, Speech and Signal Processing \textup{(ICASSP)}},
  pages={7009--7013},
  year={2020},
}

@article{robustasr4,
  title={Attention-based latent features for jointly trained end-to-end automatic speech recognition with modified speech enhancement},
  author={Yang, Da-Hee and Chang, Joon-Hyuk},
  journal={Journal of King Saud University-Computer and Information Sciences},
  volume={35},
  number={3},
  pages={202--210},
  year={2023},
  publisher={Elsevier}
}

@inproceedings{FM,
  title={Flow Matching for Generative Modeling},
  author={Yaron Lipman and Ricky T. Q. Chen and Heli Ben-Hamu and Maximilian Nickel and Matt Le},
  booktitle={Proc. International Conference on Learning Representations \textup{(ICLR)}},
  year={2023},
}

@inproceedings{scoreM,
  title={Score-based generative modeling through stochastic differential equations},
  author={Song, Y. and Sohl-Dickstein, J. and Kingma, D. P. and Kumar, A. and Ermon, S. and Poole, B.},
  booktitle={Proc. International Conference on Learning Representations \textup{(ICLR)}},
  year={2021}
}

@article{flowplc,
  title={Flow-PLC: Towards efficient packet loss concealment with flow matching},
  author={Yang, Da-Hee and Chang, Joon-Hyuk},
  journal={IEEE Signal Processing Letters},
  year={2025},
  publisher={IEEE}
}

@inproceedings{flowsep,
  title={Flowsep: Language-queried sound separation with rectified flow matching},
  author={Yuan, Yi and Liu, Xubo and Liu, Haohe and Plumbley, Mark D and Wang, Wenwu},
  booktitle={Proc. IEEE International Conference on Acoustics, Speech and Signal Processing \textup{(ICASSP)}},
  pages={1--5},
  year={2025}
}

@article{dmellm,
  title={Tokenized Generative Speech Enhancement With Language Model and Flow Matching},
  author={Yang, Da-Hee and Lee, Jaeuk and Chang, Joon-Hyuk},
  journal={IEEE Signal Processing Letters},
  year={2025},
  publisher={IEEE}
}

@article{wsjdata,
  title={{CSR-II (WSJ1)} complete},
  author={Linguistic Data Consortium and others},
  journal={Linguistic Data Consortium, Philadelphia, vol. LDC94S13A},
  year={1994}
}

@inproceedings{adam,
  title={Adam: A method for stochastic optimization},
  author={Kingma, Diederik P and Ba, Jimmy},
  booktitle={Proc. International Conference on Learning Representations \textup{(ICLR)}},
  year={2015}
}

@article{silu,
  title={Sigmoid-weighted linear units for neural network function approximation in reinforcement learning},
  author={Elfwing, Stefan and Uchibe, Eiji and Doya, Kenji},
  journal={Neural networks},
  volume={107},
  pages={3--11},
  year={2018},
  publisher={Elsevier}
}

@incollection{chime,
  title={The \textup{CHiME} challenges: Robust speech recognition in everyday environments},
  author={Barker, Jon P and Marxer, Ricard and Vincent, Emmanuel and Watanabe, Shinji},
  booktitle={New era for robust speech recognition: Exploiting deep learning},
  pages={327--344},
  year={2017},
  publisher={Springer}
}

@inproceedings{dnsmos,
  title={\textup{DNSMOS}: A non-intrusive perceptual objective speech quality metric to evaluate noise suppressors},
  author={Reddy, Chandan KA and Gopal, Vishak and Cutler, Ross},
  booktitle={Proc. IEEE International Conference on Acoustics, Speech and Signal Processing \textup{(ICASSP)}},
  pages={6493--6497},
  year={2021}
}

\end{document}